# A testing facility for AO on-sky demonstrations at the Copernico's Telescope within the ADONI framework

S.Chinellato [a,b,1], R.Ragazzoni [a,b], J.Farinato [a,b], D.Greggio [a,b], M.Dima [a,b], F.Biondi [a,b], S.Benetti [a], M.Bergomi [a,b], E.Cappellaro [a], E.Carolo [a], V.Chiomento [a], A.Frigo [a], L.Lessio [a,b], D.Magrin [a,b], L.Marafatto [a,b], G.Martorana [a], E.Portaluri [a,b], M.Rebeschini [a], L.Tomasella [a], L.Traverso [a], M.Turatto [a], G.Umbriaco [a,b], D.Vassallo [a,b,c], V.Viotto [a,b]

[a]INAF, Astronomical Observatory of Padova, Vicolo dell'Osservatorio, 5 35122 Padova, Italy;
[b]ADONI, Laboratorio Nazionale di Ottica Adattiva, Italy;
[c]Dipartimento di Fisica e Astronomia, Università degli Studi di Padova, Italy

## ABSTRACT

In the context of ADONI - the ADaptive Optics National laboratory of INAF – we are arranging for a facility, accessible to the AO community, in which visiting multi-purpose instrumentation, e.g. systems and prototypes of innovative AO concepts, may be directly tested on sky. The facility is located at the 182cm Copernico telescope in Asiago, the largest telescope in Italy, at its Coudè focus, for which refurbishment activities are carried out, given that this focus was initially foreseen in the design, but never implemented and used till today. The facility hosts a laboratory where specialized visiting AO instrumentation may be properly accommodated on an optical bench for on-sky demonstrations. We present the current status of the facility, describing the opto-mechanical design implemented at the telescope that allows to redirect the light toward the Coudè focus, the tests on the opto-mechanics carried on for stability verification, the integration of the optical and mechanical components within the preexisting structure.

## 1. INTRODUCTION

Within the ADONI (ADaptive Optics National laboratory of Italy), we foresee to set-up a laboratory accessible to the Italian and international AO community, with the purpose of proposing a testing facility where AO experimentations can be conducted[1]. The 182cm Copernico Telescope, located in Asiago (Italy), is being refurbished in order to create a suitable place to set-up this public facility, where a common optical bench will be made available at the Coudé focus to host visiting instrumentation. Testing of critical sub-systems or components, which may be part of instruments under construction, or prototypes of innovative concepts, which may require on-sky demonstrations, is the aim of the facility.

In this paper we give an overview of the activities carried on for the implementation of the facility, describing the refurbishment at the hosting structure that will allow to operate at the telescope's Coudé focus: the optical and mechanical design is described in Section 2. After a phase of procurement of components, and their characterization, integration of mechanics within the existing structure has been completed and alignment phase is currently going on, see Section 3 for details. A description of the set-up for the laboratory in the dedicated area is reported in Section 4.

## 2. THE COUDÈ FOCUS AT THE COPERNICO TELESCOPE

Among INAF-Astronomical Observatory of Padova (OAPd), the Copernico Telescope is the largest optical telescope in Italy, a classic Cassegrain with equatorial mount, having a primary mirror as large as 182cm diameter and with equivalent focal number f/9[2]. Located in the Asiago plateau, north-east of Italy, about one hundred km from Padova, at an altitude of 1366 m (Mt. Ekar, 11°34′08.42″ E, 45°50′54.52″ N), this site is characterized by a continental climate, with dry winters and rainy springs. Summer time (starting from August) is on average favorable for observations. The seeing is quite variable during the year, with an average ∼ 2″, but nights with seeing around 1″ are frequently registered[3]. Two main instruments are currently active at the telescope, a low resolution spectrograph and imager (AFOSC, Asiago Faint Object Spectrograph and Camera) and a high resolution spectrograph (Echelle): complementary for spectral resolution and therefore mounted in turn at the telescope following the moon cycle (AFOSC being usually available during grey and dark time). In addition, Aqueye (Asiago Quantum Eye), an ultrafast photon counter, is also frequently used.

---
[1] simonetta.chinellato@oapd.inaf.it

The realization of the facility for AO experimentation implied, from one side the refurbishment of the telescope to open the Coudé focus, that was foreseen in the initial project but never implemented till today, and the setting-up of the laboratory dedicated to the facility. In the design from the 70's the idea was that of having an exchangeable secondary mirror: by switching this and with the aid of only three folding mirrors it was foreseen to arrive down to the Coudé focal station. This is not feasible today, the telescope is active and in the current configuration used by other instrumentation. Since a key concern for these implementations was that they would not change the telescope's performance when adding new opto-mechanical components, a more suitable configuration was that of redesigning the optical train down to the new focal station.

In the newly conceived optical design, light passing through the primary and secondary mirrors, is redirected by the insertion of a 45 degree tertiary mirror towards one of the two Nasmyth arms and passes all the way through the fork down to the Coudé focal station located at a floor below the dome level. The optical design implemented from the tertiary mirror down to the new focal station is shown in Fig. 1. The new optical path consists of seven optical components, three achromatic doublets (L1, L2, L3) and four folding mirrors (F1, F2, F3, F4), all integrated in the fork, except the fourth and last mirror being outside so to redirect light on the optical table.

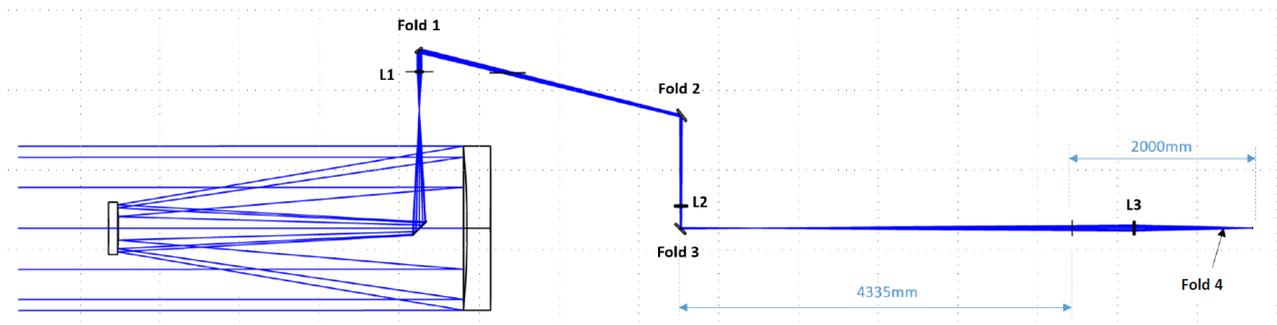

Figure 1. Optical design showing the optical train from the tertiary mirror down to the Coudé focus. Seven optical components compose the optical train: three achromatic doublets (L1, L2, L3) and four folding mirrors (Fold1, 2, 3, 4) are inserted to redirect the beam.

A first folding mirror (Fold1) redirects the beam down into the fork of the telescope, the second (Fold2) and third (Fold3) mirrors send it down to the focus located at a lower level. A list of the optical components with their dimensions is summarized in in Tab. 1. The optical components and mounts for lenses and mirrors are mostly commercial. These are assembled together and integrated along the fork, bounded within mechanical supports designed and made of the same material of the telescope's structure to minimize differential thermal expansion.

Table 1. Optical components specification's for the Coudé's optical station.

| Optical components | F (mm) | Phys. Aperture (mm) |
|---|---|---|
| **L1** | 400 | 75 |
| **L2** | 2000 | 150 |
| **L3** | 1000 | 150 |
| **Fold 1** | - | 76.2 |
| **Fold 2** | - | 203.2 |
| **Fold 3** | - | 203.2 |
| **Fold 4** | - | 100 |

With the insertion of three achromatic doublets (L1, L2, L3), this configuration allows to have a beam at the Coudé focus with main specifications summarized in Tab. 2. One of the main characteristics of this design is the beam's telecentricity, giving a great versatility to the testing facility and allowing for a large range of optical set-ups to be arranged for experimental testing.

Table 2. Coudé focus optical specifications.

| Specifications | |
|---|---|
| Telescope diameter [mm] | 1820 |
| Coudé F/# | 20 |
| Plate scale [arcsec/mm] | 5.667 |
| Radial FoV [arcmin] | 1.2 |
| Wavelength range [nm] | Visible (400 - 1000) |

Mechanical supports have been designed in correspondence of the optics, as shown in the sketch of the mechanical design in Fig. 2 below. The path either follows the fork or passes on the outside through apertures already present in the mechanical structure of the fork. A key concern for the implementations was that these would not change the telescope performance when placing new opto-mechanical components. As far as possible, supports fit into the structure and are fixed in correspondence of pre-existing interfaces inside the fork, simplifying in this way the integration of all the components.

The first assembly – consisting of lens L1 and mirror Fold1 - is designed to be in correspondence of the Nasmyth 2 platform, just outside the fork and supported on a separate external structure in order to have a good accessibility during integration. The folding mirror Fold2 and the assembly – consisting of Lens L2 and mirror Fold3 – are kept together by a structure and bounded to the internal part of the fork, where connecting flanges are already present. The third and last optical element – lens L3 – is supported by an external structure, that is attached at the end side of the telescope tube to the base not subject to its motion.

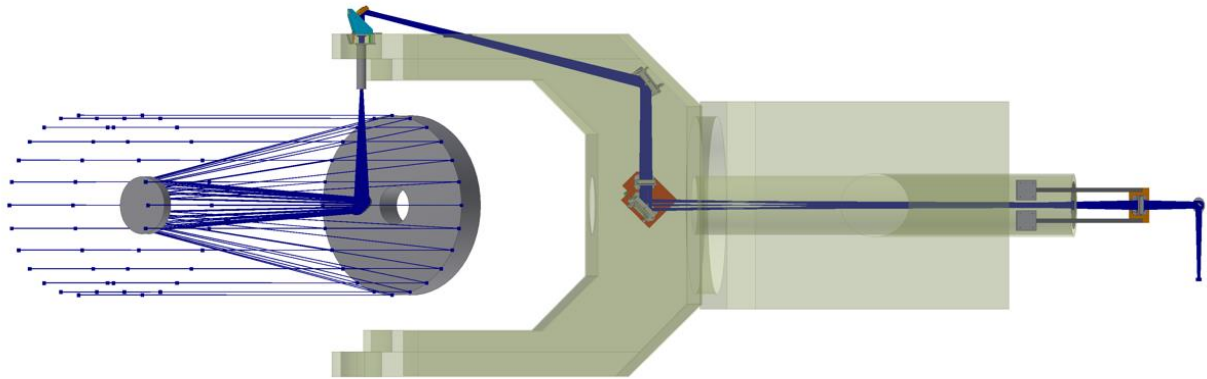

Figure 2. Mechanical design from the tertiary mirror down to the Coudé focal station.

## 3. CHARACTERIZATION, INTEGRATION AND ALIGNMENT

The Coudé focal station is characterized by being gravity invariant, the redirected beam arriving to this focus passes through all the opto-mechanical components inside the fork that are instead subject to the telescope's motion, hence they undergo gravitational forces. Before integrating the optical train in the overall system, tests on opto-mechanical components, i.e. lenses, mirrors, and mounts, have been conducted in laboratory to evaluate stability. For all components, we proceeded with the mechanical characterization of optical mounts to evaluate the flexures they experience. That is, mounts, that range in dimension from 76.5 to 203.2mm, underwent maximum tensions to retrieve an upper limit for the expected flexures.

After each mount has been characterized, flexures measured are fed to optical simulations to recover the maximum total displacement of the optics in correspondence of the tilt, for all elements and in all directions (worst case). In some cases, components showed critical values for the measured displacements and were reinforced where necessary to reduce flexures. When re-simulated, reduced radial displacements were measured. The maximum radial displacement on the focal plane is recovered and results are shown in Tab. 3, the third lens and the fourth mirror are not being considered since they are not bounded to the rotating structure, but fixed to the base of the telescope.

Table 3. Maximum radial displacement on the focal plane.

| Optical Component | Diameter (mm) | Max radial displacement on focal plane (") |
|---|---|---|
| L1 – Lens | 75 | 0.03 |
| F1 – Mirror | 150 | 2.77 |
| F2 – Mirror | 150 | 4.64 |
| L2 – Lens | 76.2 | 0.014 |
| F3 - Mirror | 203.2 | 2.93 |
| L3 - Lens | 203.2 | NA |
| F4 - Mirror | 100 | NA |

The total radial displacement on focal plane recovered from a Monte Carlo simulation (with 10000 iterations) is of the order of 7.6", corresponding to a FoV of 10%.

After reinforcement resulted in having acceptable values, all mounts and corresponding mechanics were integrated within the telescope's structure. Alignment phase in currently being carried on, using a laser to define the optical axis, firstly dealing with the alignment of the mirrors and successively with lenses. Analyzing measurements of the flexures of the newly inserted mechanical supports, due that they are subject to the telescope's motion, and comparing them with the simulated flexures deriving from the test on the mounts, led to the breakdown of components, allowing the separation of the contribution of flexures due to the mounts from those ascribed to mechanical supports. This entailed the reinforcement of the mechanical supports, where necessary, to reduce flexures.

## 4. THE TESTING FACILITY

The foreseen testing facility is intended to be a permanent laboratory located at the Coudé focus. This focal station is located at a lower floor from the dome, at the same level of the control room and it is physically separated from this room. Characterized by being gravity invariant, it is useful for instruments requiring mechanical stability. Designed around the constraint of locating the optical workbench referenced to the Coudé optical axis, so that visiting instruments can be housed and experiments conducted, provides a powerful scientific and technical test bench for new instrumental concepts, which may eventually be incorporated later in the next generation telescopes.

The dedicated area has been refurbished to create the laboratory, a room of c.a. 20m$^2$ hosts the laboratory where a large optical bench (2400 x 1200 mm) is placed, on which users can set up their own instruments and AO testing will be carried on. Indeed, the laboratory is enclosed in a light-tight room that can be darkened as needed; also a key requirement is to keep a stable temperature control, it is in fact also thermally insulated from the surrounding environment. In Fig. 3 pictures of the laboratory are shown. The optical axis exiting from the telescope's structure is deviated by a folding mirror (Fold 4) that redirects the beam allowing the focus to fall on the optical bench.

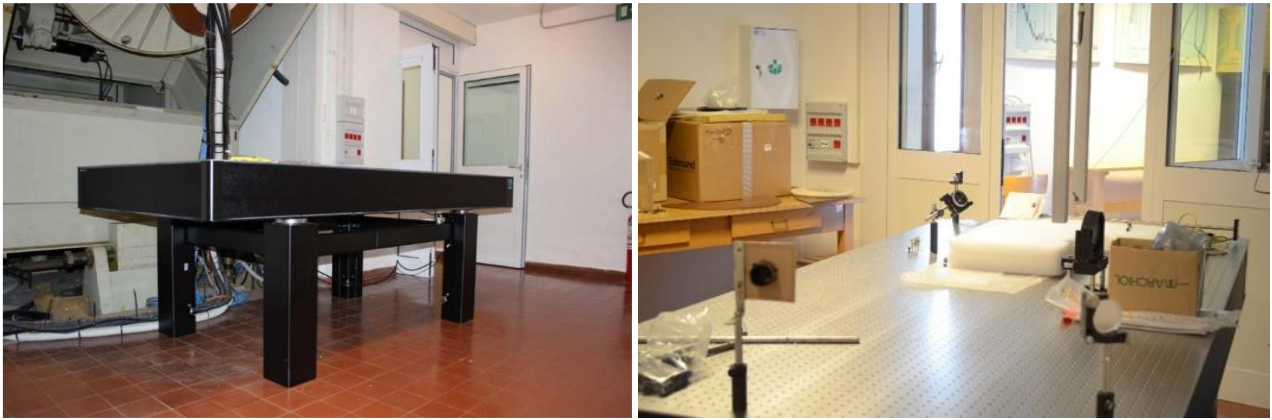

Figure 3. A view of the laboratory at the Coudé focal station of the Copernico Telescope as it has been refurbished: a light tight room with new thermal isolation and electronics. An optical table has been positioned as workbench to host visiting instruments.

This facility will be of particular interest for on-sky experiments in high angular resolution imaging, for the implementation of novel concepts of wavefront sensing, the comparison of different kinds or concepts of wavefront sensors, the adoption of specialized pupil planes aperture (e.g. double opening or multiple apertures), and any new technique to be tested. Among the existing laboratories that have access to on-sky testing, we recall GHRIL[4], the Ground-Based High-Resolution Imaging Laboratory, on the 4.2m William Herschel Telescope (WHT): a facility that presents an enclosure with an optical bench allowing access to one of its Nasmyth foci. For instance, the Pseudo Infinite Guide Star (PIGS)[5] experiment related to the implementation of a novel concept of wavefront sensing, was directly verified on-sky at the WHT. Once proved its functionality in a laboratory experiment, the entire sensor was then tested under more realistic conditions on sky. A public facility for direct demonstrations on-sky is the Visitor Focus at VLT that permits innovative observations by instrument teams using their own stand-alone instruments: the implementation of techniques in high angular resolution imaging has been tested here, as for the APE experiment[6], for example. APE was used by ESO to gain experience in controlling segmented primary mirrors in preparation for E-ELT, the experiment tested various phasing techniques and explored their advantages and limitations. Testing the adoption of specialized pupil planes aperture to be implemented in telescopes of new generation is another example of possible experimental set-up. Interferometric imaging for the LBT testing, in fact, was carried on at the 1.5 m TIRGO IR telescope, to simulate the pupil of the LBT telescope in order to record realistic LBT-like data[7].

## 5. CONCLUSIONS

The public facility dedicated to AO experimentation is being developed at the Copernico Telescope in Asiago and will allow for direct on-sky testing of newly conceived technologies. We have shown the refurbishment activities that have been carried on to open the Coudé focal station where the laboratory will be hosted. The newly opto-mechanical design that is implemented is described, along with the characterization of the components, which aided in their integration, and currently the alignment phase is going on. By simply rotating of the tertiary mirror, it is possible to select the Coudé focal station, this arrangement allows experiments to be carried out by simply mounting systems to be tested on the optical bench. The facility will have the advantage of having great versatility due to its telecentricity, allowing for AO multi-purpose visiting instrumentation.